\documentclass[letterpaper,twocolumn,english,aps,prl,reprint,floatfix,showpacs,tightenlines,superscriptaddress]{revtex4-1}
\usepackage[]{fontenc}
\usepackage[latin9]{inputenc}
\usepackage{amsmath}
\usepackage{graphicx}
\usepackage{amssymb}
\usepackage{esint}

\makeatletter
\@ifundefined{textcolor}{}
{%
 \definecolor{BLACK}{gray}{0}
 \definecolor{WHITE}{gray}{1}
 \definecolor{RED}{rgb}{1,0,0}
 \definecolor{GREEN}{rgb}{0,1,0}
 \definecolor{BLUE}{rgb}{0,0,1}
 \definecolor{CYAN}{cmyk}{1,0,0,0}
 \definecolor{MAGENTA}{cmyk}{0,1,0,0}
 \definecolor{YELLOW}{cmyk}{0,0,1,0}
 }

\usepackage{hyperref}

\newcommand{\tr}{\mathrm{tr}}

\newcommand{\1}{\leavevmode{\rm 1\ifmmode\mkern  -4.8mu\else\kern -.3em\fi I}}

\makeatother

\usepackage{babel}

\makeatother

\usepackage{babel}

\begin{document}

\title{Exact infinite-time statistics of the Loschmidt echo for a quantum
quench}

\author{Lorenzo Campos Venuti}

\affiliation{Institute for Scientific Interchange (ISI), Viale S. Severo 65, I-10133
Torino, Italy }

\author{N.~Tobias Jacobson}

\affiliation{Department of Physics and Astronomy and Center for Quantum Information
Science \& Technology, University of Southern California, Los Angeles,
California 90089-0484, USA}

\author{Siddhartha Santra}

\affiliation{Department of Physics and Astronomy and Center for Quantum Information
Science \& Technology, University of Southern California, Los Angeles,
California 90089-0484, USA}

\author{Paolo Zanardi}

\affiliation{Department of Physics and Astronomy and Center for Quantum Information
Science \& Technology, University of Southern California, Los Angeles,
California 90089-0484, USA}

\affiliation{Institute for Scientific Interchange (ISI), Viale S. Severo 65, I-10133
Torino, Italy }
\begin{abstract}
The equilibration dynamics of a closed quantum system is encoded in
the long-time distribution function of generic observables. In this
paper we consider the Loschmidt echo generalized to finite temperature,
and show that we can obtain an exact expression for its long-time
distribution for a closed system described by a quantum XY chain following
a sudden quench. In the thermodynamic limit the logarithm of the Loschmidt
echo becomes normally distributed, whereas for small quenches in the
opposite, quasi-critical regime, the distribution function acquires
a universal double-peaked form indicating poor equilibration. These
findings, obtained by a central limit theorem-type result, extend
to completely general models in the small-quench regime.
\end{abstract}

\pacs{03.65.Yz, 05.30.-d}

\maketitle

\paragraph*{Introduction}

Imagine an isolated quantum system, say the laboratory, prepared in
a state $\rho_{0}$. According to the laws of quantum mechanics, the
state will evolve unitarily into $\rho\left(t\right)$. The average
result of a measurement of an observable $O$ will be the time average
$\overline{\langle O\left(t\right)\rangle}:=T^{-1}\int_{0}^{T}\langle O\left(t\right)\rangle dt$,
where $T$ is the measurement time. Since $T$ is much larger than
the microscopic time scales of the system it is often set to infinity
for mathematical clarity. Now, the postulates of statistical mechanics
assert that the time-averaged expectation value is indistinguishable
from that obtained using the statistical microcanonical ensemble.
Although this postulate is confirmed by a number of numerical simulations
(see e.g.~\cite{rigol08}), to date no explanation exists for why
this is so. In other words, the mechanisms of thermalization in quantum
systems are unknown (though there exist possible approaches such as
the eigenstate thermalization hypothesis \cite{deutsch91,*srednicki94}
or normal typicality \cite{vonneumann1929,*goldstein2009,*tasaki2010}).

In such a context it is important to have exact results, at least
for some particular cases, which can serve to guide our intuition.
Ideally one is interested in the full, long-time statistics of a generic
observable $\langle O\left(t\right)\rangle$. This article provides
a result in this direction. Namely, concentrating on the Loschmidt
echo, we will obtain its exact, long-time distribution function and
investigate the effects that proximity to critical points has on the
equilibration dynamics. In the thermodynamic limit, also called the
off-critical regime, i.e.~when the system size is much larger than
all length scales of the system, we will see that a central limit
theorem result applies leading to universal Gaussian equilibration.
In the opposite regime of quasi-criticality, where the correlation
length is equal to or larger than the system size, we will again find
universal behavior, although one in which fluctuations are large and
thermalization does not occur.

The scenario we consider here is that of a quantum quench, generalized
to the mixed case. A closed system is initialized in the state $\rho_{0}$
commuting with the Hamiltonian $H_{0}$. The system is then instantaneously
quenched and left to evolve according to Hamiltonian $H_{1}$. This
is an important generalization, since in principle there is no reason
why the {}``initial'' state of the system should be pure. In particular,
for its experimental relevance we will use Gibbs initial states $\rho_{0}\sim e^{-\beta H_{0}}$.
Such a situation is in fact often realized in the laboratory by first
thermalizing the system by putting it in contact with an external
reservoir and then detaching the reservoir.

The quantity we consider is the Loschmidt echo (LE), which generalized
to the mixed case is given by\[
\mathcal{L}\left(t\right)=F\!\left(\rho\left(t\right),\rho_{0}\right),\,\, F\!\left(\rho,\sigma\right)=\left(\tr\sqrt{\rho^{1/2}\sigma\rho^{1/2}}\right)^{2}.\]
 Here $F$ is the Uhlmann fidelity \cite{uhlmann76} which characterizes
the degree of distinguishability between two mixed states. Note that
if either (or both) of $\rho$ and $\sigma$ is pure, the Uhlmann
fidelity simplifies to $F\!\left(\rho,\sigma\right)=\tr\left(\rho\sigma\right)$.

\paragraph*{The quantum XY chain}

The model we investigate here is the quantum XY chain in a transverse
magnetic field, \begin{equation}
H=-\sum_{i=1}^{L}\frac{(1+\gamma)}{2}\sigma_{i}^{x}\sigma_{i+1}^{x}+\frac{(1-\gamma)}{2}\sigma_{i}^{y}\sigma_{i+1}^{y}+h\sigma_{i}^{z}.\label{eq:XY_Hamiltonian}\end{equation}

A Jordan-Wigner transformation brings Eq.~(\ref{eq:XY_Hamiltonian})
to a quadratic form in Fermi operators $c_{i}$, and hence can be
exactly diagonalized. At zero temperature the model (\ref{eq:XY_Hamiltonian})
displays two kinds of quantum phase transition lines in the $\left(h,\gamma\right)$
plane. For $h=\pm1$ and $\gamma\neq0$ the model is in the Ising
universality class described by a $c=1/2$ conformal field theory
(CFT). Instead, in the segment $\gamma=0$, $\left|h\right|\le1$
the underlying CFT has central charge $c=1$. To specify completely
the problem we must fix boundary conditions (BC's). As is customary
\cite{barouch70}, to avoid unnecessary complications we will fix
BC's on the fermions \cite{nota-BCs}. Diagonalization brings Eq.~(\ref{eq:XY_Hamiltonian})
to free Fermion form: $H=\sum_{k}2\Lambda_{k}\eta_{k}^{\dagger}\eta_{k}$.
Our choice of BC's fixes quasimomenta to be quantized according to
$k=\left(2n+1\right)\pi/L$, $n=-L/2,\ldots,L/2-1$, whereas the single-particle
dispersion is $\Lambda_{k}=\sqrt{\left(\cos k+h\right)^{2}+\gamma^{2}\sin^{2}k}$.

\begin{figure}
\begin{centering}
\includegraphics[width=5cm]{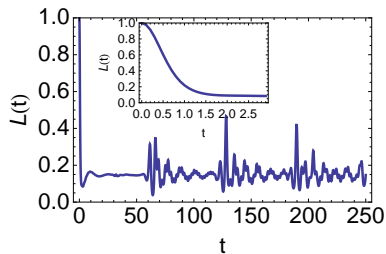} 
\par\end{centering}

\caption{Typical behavior of $\mathcal{L}(t)$. The inset shows Gaussian behavior
for short times, as happens for the pure case \cite{lcv2010a}. Here
$L=100$, $\beta=6$, $h_{0,1}=1$, $\gamma_{0}=0.5,\,\gamma_{1}=0.8$.
\label{fig:LE_behavior}}

\end{figure}

The Loschmidt echo has been shown for the XY chain to be \cite{PZ07}
$\mathcal{L}(t)=\prod_{k>0}f_{k}\left(\Lambda_{k}^{1}t\right)$, with

\begin{equation}
f_{k}\left(\Lambda_{k}^{1}t\right)=\left[\frac{1+\sqrt{c_{k}^{2}-\left(c_{k}^{2}-1\right)\alpha_{k}\sin^{2}(\Lambda_{k}^{1}t)}}{1+c_{k}}\right]^{2}\label{eq:XY_Loschmidt_echo}\end{equation}
 where $c_{k}=\cosh\left(\beta\Lambda_{k}^{0}\right)$, $\alpha_{k}=\sin^{2}(\Delta\theta_{k})$
, $\Delta\theta_{k}=\theta_{k}^{1}-\theta_{k}^{0}$ and $\theta_{k}=\arctan\left[\gamma\sin\left(k\right)/\left(h+\cos\left(k\right)\right)\right]$.
From its explicit form we can read off a number of important points
which we will use extensively in the following: i) the time-dependence
is governed by $L/2$ frequencies $\Lambda_{k}^{1}$, ii) the LE is
a product of an extensive number of terms, and in particular iii)
the LE is a product of $L/2$ functions over the $L/2$ allowed values
of $k$. The dependence on $k$ is analytic everywhere except for
the critical points ($\gamma=0$ and $\left|h\right|\le1$ or $\left|h\right|=1$
and $\gamma\neq0$). No singularity other than those expected at criticality
emerges. 

Typical behavior of $\mathcal{L}(t)$ is depicted in Fig.~\ref{fig:LE_behavior}.
The LE quickly drops from unity at $t=0$ and then oscillates about
its average value, with almost periodic revivals \cite{alioscia_revival}.

Following the spirit of Refs~\cite{lcv2010a,lcv2010b}, we are interested
in the distribution function of the LE seen as a random variable over
infinite time equipped with the uniform measure. The probability density
of the LE can be written as $P_{\mathcal{L}}\left(x\right):=\overline{\delta\left(\mathcal{L}\left(t\right)-x\right)}$,
where the bar denotes the time average (i.e.~$\overline{f}=\lim_{T\to\infty}T^{-1}\int_{0}^{T}f\left(t\right)dt$).
Saying that the LE spends most of the time close to a certain value
corresponds to a concentration result for $P_{\mathcal{L}}\left(x\right)$.

The moments of the LE can be computed using the methods developed
in \cite{lcv2010a}. Here one has the additional complication given
by the presence of the square-root in Eq.~(\ref{eq:XY_Loschmidt_echo}),
which must first be expanded into an infinite series. The result for
the first moment is $\overline{\mathcal{L}}=\prod_{k>0}f_{k}^{1}$,
with $f_{k}^{1}=1-\left(1-c_{k}^{-1}\right)\frac{\alpha_{k}}{2}+\frac{2c_{k}}{\left(1+c_{k}\right)^{2}}\left[\frac{2}{\pi}E\left(b_{k}\right)+b_{k}/4-1\right]$.
Here, $b_{k}=\left(1-c_{k}^{-2}\right)\sin^{2}\left(\Delta\theta_{k}\right)$
and $E$ is the complete elliptic integral of the second kind. Expanding
$f_{k}^{1}$ in the small quench regime, that is up to second order
in $\Delta\theta_{k}$, one is able to relate the dynamical quantity
$\overline{\mathcal{L}}$ to a static quantity. Specifically, one
obtains $\overline{\mathcal{L}}\simeq F\left(\rho_{0},\rho_{1}\right)^{2}$,
where $\rho_{0,1}$ are Gibbs states with Hamiltonians $H_{0,1}$.
This result extends the pure state result $\overline{\mathcal{L}}=\tr\left(\overline{\rho}^{2}\right)\simeq\left|\langle\psi_{0}|\psi_{1}\rangle\right|^{4}$
which can be recovered sending $\beta\to\infty$ \cite{rossini_base07,*rossini07}.

The distribution function for the LE in the Ising model (i.e.~$\gamma=1$)
at zero temperature was considered in \cite{lcv2010a}. Through numerical
simulations it was argued that, in the off-critical regime, two different
behaviors were observed. The distribution of the LE was seen as similar
to an exponential one, ($P_{\mathcal{L}}\left(x\right)\simeq\vartheta\left(x\right)e^{-x/\overline{\mathcal{L}}}/\overline{\mathcal{L}}$)
or to a bell-shaped Gaussian-looking one. In the next section we will
unify both of these conjectured results.

\paragraph*{Off-critical regime and Gaussian equilibration}

The form of the LE suggests that the LE should be thought of as a
product of variables. Let us then consider the new variable $Z=\ln\mathcal{L}$.
We will show that, under a very mild hypothesis, the variable $Z$
satisfies the standard central limit theorem (CLT). In particular,
in the off-critical regime, as $L\to\infty$, the rescaled variable
$Y=\left(Z-\overline{Z}\right)/\sqrt{L}$ will tend in distribution
to a Gaussian with zero mean and well-defined variance. To this aim
we will show that all the cumulants of $Z$ scale extensively, so
that for the rescaled variable $Y$ we will get $\kappa_{n}\left(Y\right)\propto L^{1-n/2}$
for $n\ge2$ while $\kappa_{1}\left(Y\right)=0$ by construction.
Hence only the first two cumulants of $Y$ survive in the $L\to\infty$
limit, thus showing Gaussianity of $Y$. In turn, Gaussianity of $Y$
implies that the LE is approximately Log-Normally distributed. This
explains the behavior observed in \cite{lcv2010a}, as a Log-Normal
has regimes where it looks approximately exponential or Gaussian.

In order to prove our assertion we need the (logarithm of the) moment
generating function of $Z$, $M^{Z}\left(\lambda\right):=\overline{e^{\lambda Z}}=\overline{\mathcal{L}^{\lambda}}$.
At this point we make the reasonable assumptions that the $L/2$ frequencies
$\Lambda_{k}^{1}$ are\emph{ rationally independent} (that is, linearly
independent over the field of rational numbers). Thanks to rational
independence (RI) we can use the theorem of averages (see e.g.~\cite{arnold_methods}
on page 286) to compute the time-average of $\mathcal{L}^{\lambda}$
as a phase space average over an $L/2$-dimensional torus \cite{nota-rational_independence}.
Our numerical simulations show that a possible rational dependence
is very mild and it would be quite unlucky to produce enough correlations
to invalidate the CLT. With RI, we obtain \[
M^{Z}\left(\lambda\right)=\prod_{k>0}g_{k}\left(\lambda\right),\,\, g_{k}\left(\lambda\right)=\frac{1}{2\pi}\int_{0}^{2\pi}\left[f_{k}\left(\vartheta\right)\right]^{\lambda}d\vartheta.\]
 Hence $M^{Z}\left(\lambda\right)=\exp\sum_{k>0}\ln g_{k}\left(\lambda\right)$.
The last steps of the proof come from the fact that $\ln g_{k}\left(\lambda\right)$
as a function of $k$ is Riemann integrable, with a finite integral,
provided we are away from critical points. Moreover, in the same region
of parameters, $\ln g_{k}\left(\lambda\right)$ (and so its integral
over $k$) is analytic in $\lambda$. Specifically, for large $L$,
we obtain, $\ln\left[M^{Z}\left(\lambda\right)\right]\simeq LG\left(\lambda\right)$,
with $G\left(\lambda\right)=\int_{0}^{\pi}\ln g_{k}\left(\lambda\right)dk/\left(2\pi\right)$
analytic in $\lambda$. Differentiating with respect to $\lambda$
we obtain that all the cumulants of $Z$ are extensive, which completes
the proof. $\square$

In particular, one has the CLT anywhere away from the critical points:
no other source of singularity emerges other than those expected at
criticality.

Let us now pause for a moment and discuss how the CLT could be violated.
One possibility is that the variance of $Z$ may grow with $L$ more
than extensively, i.e.~$\kappa_{2}\left(Z\right)\propto L^{Q}$,
with $Q>1$. This would imply that the variance of the rescaled variable
would diverge as $L\to\infty$, thus breaking the CLT. It can be shown
that $\kappa_{2}\left(Z\right)=\sum_{k>0}\kappa_{2}\left(k\right)$
with $\kappa_{2}\left(k\right)=m_{2}\left(k\right)-\left[m_{1}\left(k\right)\right]^{2}$,
and $m_{n}\left(k\right)=\frac{1}{2\pi}\int_{0}^{2\pi}\left[\ln\left(f_{k}\left(\vartheta\right)\right)\right]^{n}d\vartheta$
with $n=1,2$. By direct inspection of the integrals it turns out
that $\kappa_{2}\left(k\right)$ is a bounded function in the \emph{entire}
parameter range. Hence $\kappa_{2}\left(Z\right)\le\mathrm{const.}\times L$
\emph{also at critical points}. 

\begin{figure}
\begin{centering}
\includegraphics[width=4cm]{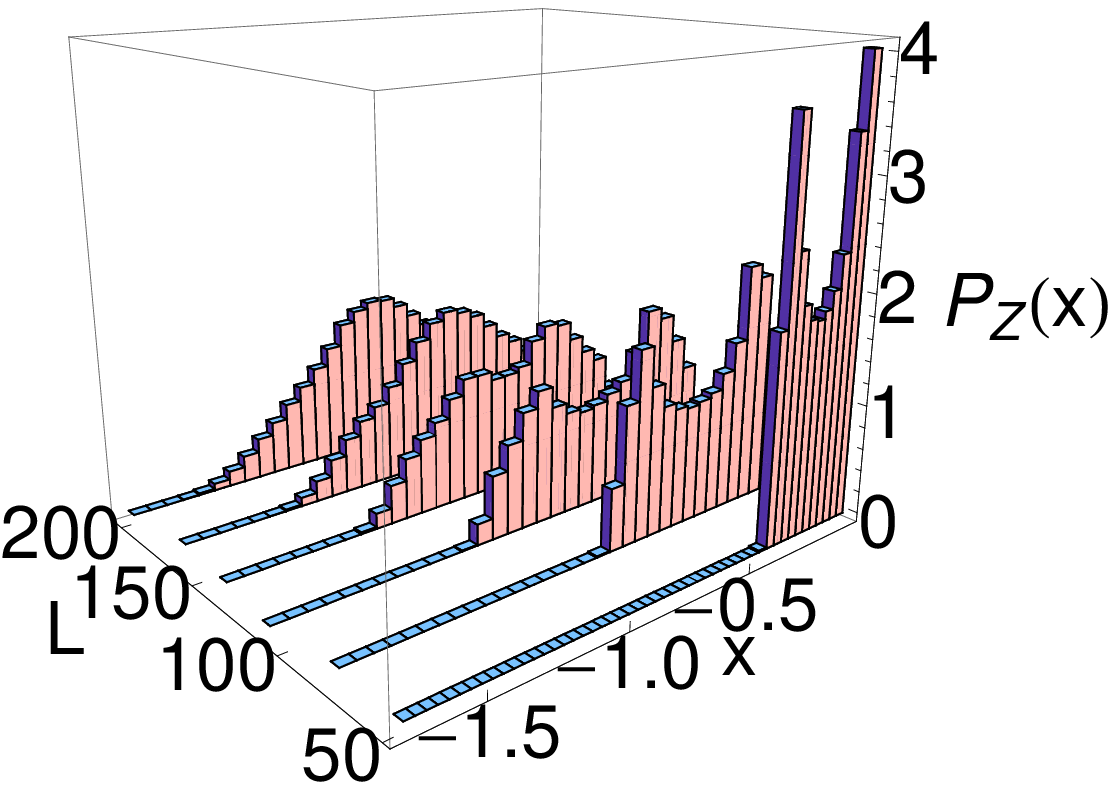}\includegraphics[width=4cm]{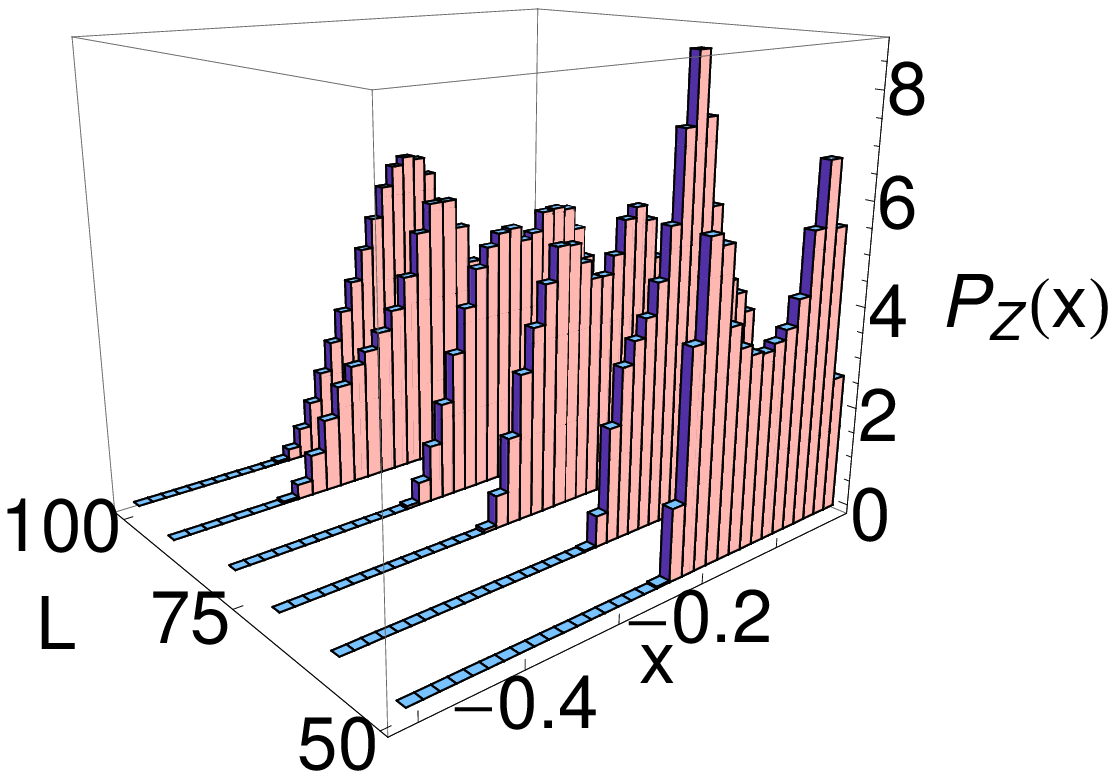} 
\par\end{centering}

\caption{$P_{Z}$ close to the Ising (left) and anisotropy transition (right).
As $L$ grows we enter the off-critical regime and $P_{Z}$ becomes
Gaussian. Close to the quasi-critical regime (small $L$) the distribution
becomes a broad, generally double-peaked function. For the anisotropy
transition, one can have $L$ for which the highest amplitudes are
nearly equal (see text). This results in a collapse from two peaks
to one. Parameters are $\beta=40$ and (left) $h_{0}=0.98,\, h_{1}=1.02,\gamma_{0,1}=1.0$
and $L=50$ to $200$ in steps of $30$, (right) $h_{0,1}=0.5,\,\gamma_{0}=0.01,\gamma_{1}=-0.01$
and $L=50$ to $100$ in steps of $10$. Another way to enter the
off-critical regime is to increase the temperature. Similar plots
are obtained replacing $L$ with the temperature $T$. \label{fig:3D_plots}}

\end{figure}

\paragraph*{Quasi-critical regime and universal critical equilibration}

In Ref.~\cite{lcv2010b} it was argued that for a small quench close
to a critical point, no observable (except for trivial constants of
motion) thermalizes. Here we will show that this result generalizes
to the mixed case considered here. Moreover, as we will see, some
universal features of the underlying critical theory show up in the
long-time distribution function. For the reasons explained above,
the right quantity to look at is the Log of the LE.

Since we are interested in the small quench regime, we expand the
Log of the LE up to the first non-zero order in $\Delta\theta_{k}$.
The constant terms add up to contribute to the average and, dropping
fourth-order terms and going to the energy variable $\omega_{j}=2\Lambda_{k_{j}}^{1}$
we arrive at \begin{equation}
\ln\mathcal{L}\left(t\right)=\overline{Z}+\sum_{j}a_{j}\cos\left(t\omega_{j}\right),\label{eq:Z-small_quench}\end{equation}
where the amplitudes are defined via $a\left(k\right)=\left(1-c_{k}^{-1}\right)\left(\Delta\theta_{k}\right)^{2}$
and $a_{j}=a\left(k_{j}\right)$.

Now we make the important observation that the quantity (\ref{eq:Z-small_quench})
is in fact a sum of $L/2$ \emph{independent random variables}. This
can be shown assuming again RI of the frequencies $\omega_{j}$. Using
the ergodic theorem one realizes that the moment generating function
of $\ln\mathcal{L}$ is simply the product of $L/2$ generating functions.
Taking the Fourier transform, one sees that each variable is distributed
according to $P_{j}\left(x\right)=\pi^{-1}\vartheta\left(a_{j}^{2}-x^{2}\right)/\sqrt{a_{j}^{2}-x^{2}}$,
with zero mean and variance $a_{j}^{2}/2$.

We are now in a position to understand what can happen at criticality
and in which sense we can expect violation of the CLT. As explained
above, the total variance, which in the small quench regime reads
$\kappa_{2}\left(Z\right)=\left(1/2\right)\sum_{j}a_{j}^{2}$, cannot
grow more than extensively. But the other extreme is possible, namely
the variances $a_{j}^{2}/2$ can go to zero as $L$ increases, and
this can happen for most of the $L/2$ variables. When this is the
case, Eq.~(\ref{eq:Z-small_quench}) effectively represents a sum
of \emph{very few} independent variables, and the CLT regime cannot
be reached. Here we notice that for an infinitesimal quench $\kappa_{2}$
is related to the fidelity susceptibility, a central object in the
so-called fidelity approach to quantum phase transitions \cite{zanardi2006,*zanardi2007,campos2007}. 

As we will see, close to criticality $a_{j}$ is a rapidly-decreasing
function of $j$, so that only few amplitudes are appreciably different
from zero. In this situation, a good approximation to the distribution
function for $Z$ is given retaining the $n_{\mathrm{max}}$ largest
amplitudes $a_{j}$ in Eq.~(\ref{eq:Z-small_quench}). Choosing $n_{\mathrm{max}}=1$,
the distribution is the just-encountered $P_{j_{\mathrm{max}}}\left(x\right)$
with square-root singularities at $\pm a_{j_{\mathrm{max}}}$. With
$n_{\mathrm{max}}=2$ the distribution is still a very spread double-peaked
one, with logarithmic singularities at $\overline{Z}\pm\left|\left|a_{1}\right|-\left|a_{2}\right|\right|$
as shown in \cite{lcv2010a}. Using the ergodic theorem it can be
shown that this distribution is precisely the density of states (DOS)
of a tight-binding model in two dimensions, with anisotropic couplings.
In general, the distribution function obtained by keeping $n_{\mathrm{max}}$
amplitudes is the density of states of a hypercubic $n_{\mathrm{max}}$-dimensional
tight binding model with anisotropic couplings $a_{j}/2$ ($j=1,\ldots,n_{\mathrm{max}}$)
in each direction. Adding more and more amplitudes, eventually the
CLT sets in and the distribution approaches a single-peaked Gaussian.
Clearly, when $n_{\mathrm{max}}$ is small the distribution function
is very spread with a large variance, so thermalization does not take
place.

Let us now discuss the behavior of $a_{j}$ close to criticality.
The XY model has two different kinds of critical regimes characterized
by different underlying effective field theories. We now consider
separately both critical regimes. First of all, note that increasing
the temperature simply has the effect of multiplying $a\left(k,T=0\right)$
by a factor $\left(1-\cosh^{-1}\left(\Lambda_{k}/T\right)\right)\le1$.
At the Ising transition we observe a large peak in $a\left(k\right)$
close to $k=\pi$. The reason for the peak has to be ascribed to the
single-particle energy vanishing as $\omega=v\left(k-\pi\right)$
(where $v=2\left|\gamma\right|$ is a velocity). The precise mechanism
has been explained in \cite{lcv2010b} for the pure case. At finite
size the quasimomenta $k$ take only discrete values. Correspondingly,
most of the weight is absorbed by those $k$'s which fall in the peak.
Other amplitudes $a\left(k_{j}\right)$ are considerably smaller.
As a result, a good approximation to the distribution can be given
by a 2D DOS as shown in Fig. \ref{fig:3D_plots}, left panel.

The situation at the anisotropy transition ($c=1$ line) is very similar,
with some notable difference due to the precise character of the $c=1$
CFT. As can be easily seen, $a\left(k\right)$ now has \emph{two }peaks,
due to the presence of two chiral (Majorana) Fermions corresponding
to the two branches of $\omega=v\left|k-k_{F}\right|$. The double-peaked
form of $a\left(k\right)$ has some detectable consequence on the
structure of the distribution function. Namely, according to different
quantization of quasimomenta (and damping factor due to temperature)
the allowed values of $k$ can fall symmetrically displaced among
the peaks. When this is the case we will observe, somehow accidentally,
a distribution function given the 2D DOS with $a_{1}=a_{2}$. In this
case the two peaks of the distribution merge into a single one, as
can be seen in Fig.~\ref{fig:3D_plots} right panel at $L=60,\,90$.

\paragraph*{Generalization}

We now give an argument in support of the validity \emph{in general}
of this scenario for small quenches. Let us restrict, for simplicity,
to zero temperature. Assuming a completely generic, non-degenerate
Hamiltonian $H=\sum_{n}E_{n}|n\rangle\langle n|$, the LE reads $\mathcal{L}\left(t\right)=\overline{\mathcal{L}}+2\sum_{n>m}p_{n}p_{m}\cos\left(t\left(E_{n}-E_{m}\right)\right)$,
where $p_{n}=\left|\langle n|\psi_{0}\rangle\right|^{2}$ for an initial
state $|\psi_{0}\rangle$. Consider now the logarithm of the LE and
expand it in the small quench parameter (that is in the perturbing
potential $V$, which we assume to be extensive). Up to second order
we obtain $\ln\mathcal{L}\left(t\right)=\overline{Z}+2\sum_{n>0}p_{n}\cos\left(t\left(E_{n}-E_{0}\right)\right)$,
where for a small quench $p_{n}=\left|\langle n|V|0\rangle\right|^{2}/\left(E_{n}-E_{0}\right)^{2}$.
If we now assume additionally RI for the energy gaps, we return to
the previous situation with $a_{j}=2p_{j}$, namely CLT away from
criticality, meaning Gaussian equilibration. Note that the total variance
is at most extensive: $\kappa_{2}\left(Z\right)=2\sum_{n>0}p_{n}^{2}\le2\sum_{n>0}p_{n}=2\chi$,
where $\chi$ is the fidelity susceptibility and is extensive by
the extensivity of $V$ and the assumption of non-criticality \cite{campos2007}.
In the quasi-critical regime only a few terms of the sum dominate,
thus breaking the CLT and leading to a universal, poorly equilibrating
regime.

\paragraph*{Conclusions}

In this letter we have considered the finite temperature generalization
of the Loschmidt echo (LE) after a quantum quench. We have proved,
under a very mild hypothesis, that away from critical points the LE
is Log-Normally distributed, whereas for small quenches close to criticality
the distribution approaches that of the density of states of a $D$-dimensional
anisotropic tight binding model, where $D$ can be considered small
(e.g.~$D=1,2$). Although these results could be obtained analytically
for the XY model considered here, we conjecture that such behavior
is in fact general and not restricted to solvable models.

LCV gratefully acknowledges support from European project COQUIT under
FET-Open grant number 2333747, NTJ from an Oakley Fellowship, and
PZ from NSF grants PHY-803304, DMR-0804914.

\bibliographystyle{apsrev}
\bibliography{ref_le}

\begin{thebibliography}{21}
\bibitem[{not({\natexlab{a}})}]{nota-BCs}
\bibinfo{note}{To be specific, we will use anti-periodic BC's for the Fermi
  operators, i.e.~$c_{L+1}=-c_{1}$.}
\bibitem[{not({\natexlab{b}})}]{nota-rational_independence}
\bibinfo{note}{Given the form of the $\Lambda_{k}$ we can expect that RI holds
  for most $\gamma,h$ (i.e. except for a set of zero measure) and for some $L$.
  From Gauss' theorem on the irreducibility of cyclotomic polynomials (see
  e.g.~\cite{galois_theory_book} Chapter 12) one can derive rational
  independence of $\left\{ \cos\left(2\pi j/L\right)\right\}$ for
  $j=0,1,\ldots,\left(L-1\right)/2$ for $L$ prime. Calling $\zeta_{j}=\exp
  i\vartheta_{j}$ with $\vartheta_{j}=2\pi j/L$, Gauss' theorem asserts that
  $\sum_{j=0}^{L-1}n_{j}\zeta_{j}=0$ with $n_{j}\in\mathbb{Z}$ implies
  $n_{j}=0$ whenever $L$ is prime. Taking real and imaginary parts one obtains
  $n_{0}+\sum_{j=1}^{(L-1)/2}n_{j}^{+}\cos\left(\vartheta_{j}\right)=0$ and
  $\sum_{j=1}^{(L-1)/2}n_{j}^{-}\sin\left(\vartheta_{j}\right)=0$, with
  $n_{j}^{\pm}=n_{j}\pm n_{j+\left(L-1\right)/2}$. Since the numbers
  $n_{j}^{+}$ and $n_{j}^{-}$ are independent, the result follows. Hence one
  can expect that such RI carries over to the $\Lambda_{k}$, but it is possible
  that the functional dependence may lift the requirement that $L$ is prime.}
\end{thebibliography}

\end{document}